\documentstyle[epsfig]{mn}

\begin{document}
\def\ltsima{$\; \buildrel < \over \sim \;$}
\def\simlt{\lower.5ex\hbox{\ltsima}}
\def\gtsima{$\; \buildrel > \over \sim \;$}
\def\simgt{\lower.5ex\hbox{\gtsima}}
\def\approxgt{\mathrel{\hbox{\rlap{\lower.55ex \hbox {$\sim$}}
        \kern-.3em \raise.4ex \hbox{$>$}}}}
\def\approxlt{\mathrel{\hbox{\rlap{\lower.55ex \hbox {$\sim$}}
        \kern-.3em \raise.4ex \hbox{$<$}}}}
\def\mnras{MNRAS}
\def\apj{ApJ}
\def\aj{AJ}
\def\apjs{ApJS}
\def\aap{A\&A}
\def\apjl{ApJL}
\def\pasp{PASP}
\def\araa{ARA\&A}
\def\pasj{PASJ}
\def\nat{Nature}

\title{XMM-Newton unveils the type~2 nature of the BLRG 3C~445}

\author[Paola Grandi]
{Paola Grandi$^1$, Matteo Guainazzi$^2$, Massimo Cappi$^1$, Gabriele Ponti$^{1,3}$\\ ~ \\
$^1$Istituto di Astrofisica Spaziale e Fisica Cosmica-Bologna, INAF, 
via Gobetti 101, I-40129 Bologna \\
$^2$European Space Astronomy Center of ESA, Apartado, 50727, E-28080 
Madrid, Spain \\
   $^3$Dipartimento di Astronomia, Universit\`a di Bologna, Via
   Ranzani 1, I--40127, Bologna, Italy}

\maketitle

\begin{abstract}
We present an observation of XMM-Newton that unambiguously reveals
the ``Seyfert 2'' nature of the Broad Line Radio Galaxy 3C 445.
For the first time the soft excess of this source has been 
resolved. It consists of unobscured scattered continuum 
flux and emission lines, likely produced in a warm photoionized 
gas near the pole of an obscuring torus.
The presence of circumnuclear (likely stratified) matter is 
supported by the complex obscuration of the nuclear region.
Seventy percent of the nuclear radiation (first component) 
is indeed obscured by a column density $\sim 4\times 10^{23}$ $cm^{-2}$, 
and $30\%$ (second component) is filtered by $\sim 7\times 10^{22}$ $cm^{-2}$.
The first component is  nuclear radiation  
directly observed by transmission through the thicker regions. 
The second one is of more uncertain nature.
If the observer has a deep view into the nucleus but near 
the edge of the torus, it could be light scattered by the inner
wall of the torus and/or by photoionized gas within the Broad Line Region
observed through the thinner rim of the circumnuclear matter.

\end{abstract}
\vspace{1.0cm}

\section{Introduction}

3C 445 (z=0.05623, Eracleous \& Halpern 2004) is a powerful FRII
galaxy (P$_{178}= 3 \times 10^{25}$ W Hz$^{-1}$ sr$^{-1}$) showing a
pair of double-lobes whose radio axis are misaligned by 2$^{\circ}$
\cite{shoe00}.  It it classified as a Broad Line Radio Galaxy (BLRG)
because of its broad and intense Balmer lines (FHWM$_{H
\beta}$=3$\times10^{3}$ km s$^{-1}$) produced in a region of size
R=1.2$\times10^{16}$ cm \cite{ost76}.  The optical continuum exhibits
decreasing polarization with increasing wavelength as expected in
polarization induced by dichroic absorption (Rudy et al. 1983, Cohen
et al. 1999).  Near infrared spectroscopical observations confirms a
substantial reddening of the BLRG region by E$_{B-V} \sim 1$ mag
(Rudy \& Tokunga 1982).  The broad $H_{\alpha}$ line is also seen in
polarized light.  Its polarization angle changes by about 45$^{\circ}$
from the red to the blue wings. This suggests a BLR inside the torus,
rotating in an equatorial plane only partially seen by the
observer. Both the red and blue-shifted emission lines from the clouds
are scattered by material surrounding the BLR.  Cohen et al. (1999)
suggested that the scattering region coincides with the inner wall of
the torus.  Dust presence in 3C445 has been recently provided by ISO
and Spitzer observations (Freudling et al. 2003, Haas et al. 2005).
In particular the radio-to infrared SED of 3C445 indicates a
predominance of dust emission in the infrared region and a negligible
contribution of synchrotron photons.  The SED modeling based on
spherical radiative transfer calculations indicates an extended dust
region of 125 pc radius and a quite large visual extinction A$_v$ = 16
(Siebenmorgen et al. 2004).  The presence of cold circumnuclear matter
has been also supported by several X-ray observations, all showing a
nuclear component strongly absorbed.  The X-ray spectrum appears
complex and structured requiring more power-laws absorbed by different
column densities (Yamashita \& Inoue 1997, Sambruna et al. 1998,
Wozniak et al. 1998, Grandi Malaguti \& Fiocchi 2006).
Interestingly, the estimated column density, ($N_H\sim 5 \times 10^{22}
- 3\times 10^{23}$ cm$^{-2}$) implies a reddening of the BLR much
larger than that observed (if a standard Galactic $E_{B-V}/N_H$ ratio
is assumed). As discussed by Maiolno et al. (2001a, 2001b), this
discrepancy can be solved if a predominance of larger dust grains
(when compared to the Galactic interstellar medium) occurs in the
circumnuclear AGN clouds.

Although classified as Broad Line Radio Galaxy, 3C 445 appears to be a
quite peculiar source, different from the other BLRGs which are
generally characterized by Seyfert~1-like X-ray spectra (Grandi,
Malaguti \& Fiocchi 2006).
Finally, we mention that a Narrow Line QSO located at only $1.3'$ from
3C~445 has been discovered by XMM-Newton (Grandi et al. 2004).  Its
spectrum is extremely soft and well represented by a power law (photon
index, $\Gamma=2.5$) plus a black body component ($kT = 117$~eV)
absorbed by Galactic $N_H$. The presence of this AGN in the vicinity
of 3C~445 adds uncertainties, casting doubts on previous X-ray
analysis based on limited spatial resolution data.
In this paper we present a XMM-Newton observation, which does
unambiguously disentangle the different components of the 3C 445
spectrum.  In particular we show that the soft X-ray radiation is a
mix of emission lines and scattered continuum produced by a
photoionized gas located well beyond the torus and the BLR.  A more
coherent geometrical/physical picture of this source, more similar to
Seyfert 2 galaxies than to BLRGs, will be proposed.

\section{Spectral analysis}

XMM-Newton observed 3C~445 on December 6 2001 for 23 ks. In this
papers data of the Reflection Grating
Spectrometer [RGS; \cite{derherder01}), and of
the European Photon Imaging Cameras (EPIC-MOS,
\cite{turner01}; EPIC-pn, \cite{struder01}] are presented.
The EPIC (MOS and PN) cameras were operated in ``small window'' mode.
The analysis was performed using the SAS software (version 6.5.0).
Since no significant pile-up was 
present in the data, single and double pixel events were selected. 
After filtering out periods of high background we obtained net
exposures of about 17~ks for the two MOS camera and $\sim$12~ks for
the pn camera.
The source plus background counts were
extracted from circular regions with a radius of 42 arcsec; the
background spectra from source--free regions in the
same chip as the source. 
The response matrices were  created using the SAS commands {\tt
arfgen} and {\tt rmfgen}. 

The RGS yields high-resolution (first order resolution
600-1700~km~s$^{-1}$) spectra in the 6--35~\AA\ (0.35--2~keV)
range. The cross-dispersion slit has a 5$\arcmin$ diameter, thus
encompassing the whole optical diameter of the optical galaxy. Data
were reduced from the {\it Observation Data files}, according to
standard procedures as in Guainazzi \& Bianchi (2006). It suffices
here to say that we have used SASv6.5 \cite{gabriel03}, and the most
updated calibration file. Source spectra were generated assuming an aspect
reconstruction centered on the nominal optical coordinates of the
3C~445 nucleus. Background spectra were generated from blank fields
accumulated over the mission.

All spectral fits presented in this paper include absorption due to a
line-of-sight Galactic column density of $N_H$=5.01$\times$10$^{20}$
cm$^{-2}$ (Murphy et al. 1966).  Errors are quoted at the 90 per cent
confidence level ($\Delta \chi^2=2.7$) for one interesting parameter.
The cosmological parameters used throughout the paper are $H_0=70$
km$^{-1}$ s$^{-1}$ Mpc$^{-1}$, $\Omega_0 = 0.73$ and q$_0$= 0.

\section{The EPIC spectrum}

The X-ray spectral complexity of 3C~455 is well known from previous
studies.  Therefore we initially decided to consider only the energy
range between 2-10 keV to avoid the highly structured soft emission
while analyzing the hard X-ray continuum.
The combination of a photoelectrically absorbed power-law and of a
Gaussian emission line profile was the first model applied to the EPIC
spectra, simultaneously fitted.  The fit was formally acceptable, but 
the power law was extremely hard ($\Gamma= 1.2$)
implying, as expected, that a more complicated model is required.
BeppoSAX data (Grandi Malaguti \& Fiocchi 2006) indicated the presence
of a reflection hump ad high energies, with a relative reflection
normalization, R=1.2. However, it was impossible to significantly
constrain the value of $R$ due to the insufficient quality of the data
and to a possible contamination of the 15-100 keV spectrum by the
nearby A2440 cluster.  Hence we aimed at verifying whether a
reflection component was required by the EPIC spectra as well.  The
quality of the fit improved after the inclusion of this component in
the first model ($\chi^2/\nu=220/238$). However,
the normalization of the reflection component was unusually large
(R$\simeq$5) and a power-law photon index still hard($\Gamma \simeq
1.43$).  As a strong reflected component can mimic an absorbed
power-law, a different model was tried.  We tried to fix the relative
normalization of the reflection component to the best-fit value
measured by BeppoSAX, and added a second power-law, with same spectral
slope as that first model, screened by a different absorbing column
density.  The fit improved again ($\Delta\chi^2 = 6$ for a decrease by
1 in the number of degrees of freedom) 
The photon index is not flat
anymore ($\Gamma \simeq 1.7$), and closer to typically observed
values.  In this scenario, a column density $\simeq 4\times10^{23}$
cm$^{-2}$ obscures $\sim 70\%$ of the hard continuum, whereas the
remaining $\sim 30\%$ of the nuclear radiation is seen through less
thick material, with $N_H \simeq 8\times10^{22}$ cm$^{-2}$.  The
Equivalent Width of the neutral or mildly ionized iron K$_{\alpha}$
fluorescent emission line, $EW \simeq 160$~eV, is in agreement with
ASCA and BeppoSAX results (Sambruna et al. 1998, Grandi et al. 2006,
Dadina 2007).  As already suggested by Wozniak et al. (1998), the cold
gas covering the nuclear region is probably the source of the Fe
K$_{\alpha}$ line.

The Fe equivalent width of $\sim 160$ eV is indeed in good agreement with the
predictions of Ghisellini Haardt \& Matt (1994), who simulated the
contribution of circumnuclear (toroidal) matter 
to the X-ray spectrum of Seyferts.  
Thus, assuming a Keplerian motion of the obscuring
matter and a black hole mass of $M=1.4\times 10^8$ M$_{\odot}$
(Bettoni et al. 2003), the intrinsic width of the Fe line provides a
rough estimate of the torus location $R \sim \frac{GM}{(c
\sigma_{Fe}/E_{Fe})^2} \sim 2.3\times10^{17}$ cm.  Taking into
account the $\sigma_{Fe}$ uncertainties, we estimate that the torus distance
can not be smaller than 0.02 pc and larger than 0.7 pc. 


\begin{table}
\caption{EPIC best-fit parameters. 
normalizations $n$ are expressed in cm$^{-2}$~s$^{-1}$~keV$^{-1}$; the
normalization of the iron line is expressed in photons cm$^{-2}$~s$^{-1}$.}
 \label{tab1}      
\centering          
\begin{tabular}{l c }     
\hline\hline       
$\Gamma$             &   1.68$^{+0.04}_{-0.02}$   \\
R                    &   1.2 (fixed) \\
N$_{H,1}$            & (4$^{+3}_{-2})\times10^{23}$ cm$^{-2}$        \\
N$_{H,2}$            & (7.8$\pm0.4) \times10^{22}$ cm$^{-2}$ \\
n$_1$                & (3.2$\pm0.1) \times10^{-3}$ \\
n$_2$                & (9.4$^{+0.5}_{-0.7}) \times10^{-4}$\\ 
n$_{unobscured}$     & (9.70$\pm0.07) \times10^{-5}$ \\
$E_{Fe~~}$   (keV)   & 6.36$\pm0.04$  \\ 
$\sigma_{Fe~~}$ (eV) & 60$^{+50}_{-40}$       \\
$F_{Fe~~}$           & 2.3$^{+0.7}_{-0.6}\times10^{-5}$\\
EW (eV)              &              160$^{+50}_{-40}$ \\ 
\hline
\multicolumn{2}{l}{Unabsorbed fluxes and luminosities} \\
$L_{2-10~~keV}$ (erg~s$^{-1}$)      &     $1.2 \times10^{44}$ \\
$Flux_{2-10~~keV}$ (erg~s$^{-1}$ cm$^{-2}$)&  1.7$\times 10^{-11}$ \\
\hline\\                                
\end{tabular}
\end{table}

Extending the 2-10 keV model to lower energies reveals a highly
structured soft emission and a remarkable similarity with
``classical'' radio quiet Seyfert 2 galaxies (Matt et al. 2004).
Driven by this analogy, we then searched for emission lines in the
3C~445 soft X-ray spectrum overimposed to a unabsorbed power law
continuum.  A summary of the best-fit parameters for the hard X-ray
spectrum is shown in Table~1, while the detected lines are listed in
Table~2.  The broadband spectrum ($0.4-10$~keV) alongside the final
model ($\chi^2/\nu=258/279$) is shown in Figure~1.

\begin{table}
\caption{Emission lines detected in the Epic soft X-rays
spectrum. $E_c$ is the centroid line energy (keV) in the source rest
frame; $F$ is the line flux in units of 10$^{-6}$ photons cm$^{-2}$  s$^{-1}$.}
\label{tab3}      
\centering          
\begin{tabular}{l c c c}     
\hline\hline\    
Tentative Identification & $E_c$ & $F$ &\\
\hline                
NVII Ly-${\alpha}$ & 0.50$^{+0.02}_{-0.03}$ & 13.9$^{+0.6}_{-0.8}$ &\\
OVII He-${\alpha}$      & 0.58$^{+0.05}_{-0.05}$ & 22.6$^{+0.6}_{-0.9}$ &\\ 
FeXVII 3d-2p   & 0.83$^{+0.03}_{-0.02}$ & 6.3$^{+2.6}_{-2.5}$ & \\
SiXIII He-${\alpha}$     & 1.80$^{+0.04}_{-0.02}$ & 3.5$^{+1.1}_{-1.2}$ &\\
\hline \\ 
\end{tabular}
\end{table}

\begin{figure}
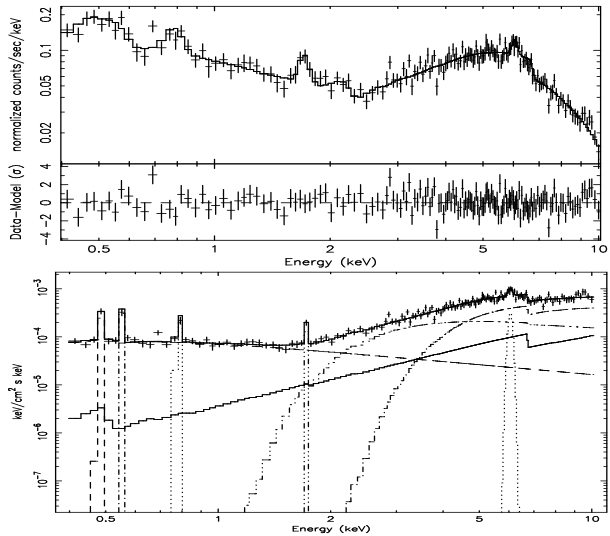

\epsfig{file=fig1l.ps,height=8cm,width=3.5cm, angle=-90}
\epsfig{file=fig1r.ps,height=8cm,width=3.5cm, angle=-90}
     \caption{3C~3445 PN spectrum ({\it upper panel}) and residuals in
term of standard deviations. ({\it lower panel}) Unfolded spectral model }
\label{fig1}
\end{figure}
The flux at 1 keV of the unabsorbed continuum is  $\sim 3\% $ of the 
direct strongly obscured primary power law.

\section{High-resolution spectroscopy in the soft X-ray band}

We were aware that the complexity of the 0.5-10 keV emission does not
exclude other possible parameterizations of the soft spectrum. For
this reason, we decided to explore the RGS spectrum in order to find
an independent support to our interpretation.
 
\begin{figure}
\epsfig{file=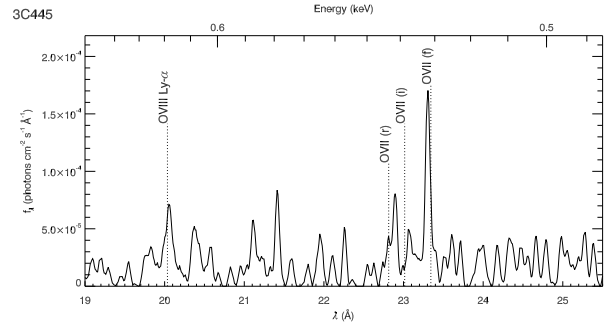,width=8.0cm}
     \caption{3C~3445 RGS spectrum of 3C~445 in the 19--25 \AA\
wavelength range. The rest-frame
wavelengths of the O{\sc vii} He-$\alpha$ triplet
and of the O{\sc viii} Ly-$\alpha$ are labeled.}
\label{fig3}
\end{figure}

To search for emission lines in the RGS spectra, ``local'' fits were
performed on 100 spectral channels wide intervals around energies were
emission lines typically observed in obscured AGN are typically found.
\cite{kinkhabwala02}.  The continuum underneath the
emission line was modeled with a $\Gamma=1$ power-law with free
normalization (given the negligible continuum level, the choice of
this value of the underlying photon index has no impact on the line
detection).  Unresolved Gaussian profiles were used to fit emission
lines.  Despite the overall low soft X-ray flux density, the forbidden
($f$) component of the O{\sc vii} triplet is detected, as well as the
O{\sc viii} Ly-${\alpha}$ (see Tab.~\ref{tab2}).
\begin{table}
\caption{List of Oxygen emission lines detected in the RGS spectrum of
3C~445 $E_c$ is the centroid line energy (keV) ; F is the line
flux  in units of 10$^{-6}$~photons cm$^{-2}$ s$^{-1}$}
\label{tab2}      
\centering          
\begin{tabular}{l c c}     
\hline\hline       
Identification & $E_c$ & F \\
\hline                    
O{\sc vii} He-${\alpha}$ (f) & $0.5619 \pm^{0.0003}_{0.008}$ & $31^{+22}_{-14}$ \\
O{\sc vii} He-${\alpha}$ (i) & $E_c (f) + 0.0077$ & $<$0.9 \\
O{\sc vii} He-${\alpha}$ (r) & $E_c (f) + 0.0130$ & $<$1.0 \\
O{\sc viii} Ly-${\alpha}$ & $0.6527 \pm^{0.0009}_{0.0010}$ & $11^{+12}_{-8}$ \\
\hline                  
\end{tabular}
\end{table}
A zoom of the RGS spectra around in the energy range where He- and H-like
Oxygen transitions are shown in Fig.~\ref{fig3}.
Interestingly enough, the best-fit centroid energy of the O{\sc vii}
H-${\alpha}$ triplet components is blue-shifted by $v = 430
\pm^{220}_{160}$~km~s$^{-1}$, suggesting that the line emitting gas is
outflowing (the typical statistical uncertainties on the RGS aspect
solutions are $\simeq$100~km~s$^{-1}$ at this wavelength). However,
this shift is not confirmed in the lower statistical quality O{\sc
viii} Ly-${\alpha}$ line.  Although it is potentially possible that
O{\sc vii} and O{\sc viii} transitions are emitted in dynamically
separated\ phases of the photoionized gas, we still consider the
evidence for outflowing gas in the nuclear environment of 3C~445 as
tentative, to be confirmed by future, more accurate measurements.

A handful of Fe-L transitions are marginally detected as well, and not
reported in Tab.~\ref{tab1}. They will not be further discussed
hereafter.

The detection of the He-like Oxygen triplet $f$ component is a clear
indication for a photoionized plasma. However, given the low
significance of the detection, quantitative plasma diagnostics based
on the triplet component intensity ratios \cite{gabriel69,porquet00}
are partly ambiguous.  Only lower limits can be derived for the $R$
($>1.7$) and $G$ ($>1.9$). Although fully consistent with photoionized
plasma, these value do not rule out collisional ionization.  Guainazzi
\& Bianchi 2007 proposed a difference diagnostics plane to
discriminate {\it on a statistical basis} photoionized- (AGN) from
collisionally ionized (starburst) powered sources: based on the
intensity of the $f$ component of the O{\sc vii} He-like triplet
(dubbed $\eta$ ratio, once normalized to the O{\sc viii}
Ly-${\alpha}$) against the integrated luminosity $L_=$ of He- and
H-like Oxygen lines.  The values derived from the RGS spectrum of
3C~445 ($\log (\eta) = 0.5 \pm^{0.2}_{0.7}$; $L_O \sim 3 \times
10^{41}$~erg~s$^{-1}$) place 3C~445 in the plane region preferentially
occupied by obscured AGN.
\section{Discussion}

\subsection{Nuclear continuum and obscuring gas}

Although 3C445 is optically classified as BLRG, its X-ray spectrum is
similar to that observed in Seyfert 2 galaxies.  The nuclear continuum
is absorbed and the soft excess is well fitted by a weak power law
plus emission lines.  The hard X-ray emission is attenuated by a
complex absorber with 70$\%$ of the nuclear continuum absorbed by a
column density of $N_{H1}\sim 4\times 10^{23}$ $cm^{-2}$ (first
component) and $\sim 30\%$ by a smaller column of $N_{H2}\sim 7\times
10^{22}$ $cm^{-2}$ (second component). About $3\%$ is attenuated
only by the Galactic $N_{H}$ (third component).  These results can be
most straightforwardly interpreted using an analogy with typical
spectra of radio-quiet obscured AGN.  A torus obscures the nuclear
continuum (first component) that is probably associated to an
accretion flow, while ionized gas, produces the soft lines and, acting
as a mirror, scatters the X-ray continuum (third component) towards
the observer beyond the outer rim of the torus.  

The origin of the second component is less obvious.
Considering that 3C445 is a powerful radio galaxy, it seems natural to
associate it to a jet.  If this is the case, the
strength of the non-thermal beamed radiation (second component) is
smaller than the accretion flow (first component) by a factor $\sim
0.3$ in the 2-10 keV band.
In spite of  the apparent agreement  with the results of Grandi $\&$ Palumbo
(2007), who have attempted a first separation between the nuclear  jet and
disk in three BLRGs (3C120, 3C382 and 3C390.3),
the hypothesis is not completely convincing.
From the radio flux ratio of the approaching and
receding jets (Leahy et al 1997), we can obtain an estimate of the jet
inclination, $i\le 60^{\circ}$ . This is a rather large value if compared
for example, with 3C120 and 3C390.3 that show inclination angle less than
14$^{\circ}$ and 33$^{\circ}$, respectively (Eracleous \& Halpern 1999).
Although a  large angle of view in 3C 445 justifies its
strong nuclear absorption, it does not justify the
intense luminosity of the putative jet.
An inclination of  $i\sim 60^{\circ}$ implies a  small Doppler
factor $\delta$\footnote{$\delta=[\gamma(1-\beta \cos
i)]^{-1}$, being $\gamma=(1-\beta^2)^{-1/2}$ and $\beta$ the bulk velocity in
unit of c} and, as a consequence a reduced
flux amplification ($F_o=\delta^4 F_i$).
If the analogy with the other BLRGs is correct, in 3C 445
the jet/disk ratio should be smaller at least by a factor ten.

Alternatively, the second component  could be 
disk radiation scattered within the torus and
seen through the thinner layers of the obscuring matter.  
If the gas distribution in the circumnuclear region indeed
resembles that expected in Seyfert~2 galaxies, we could be looking at
nuclear region through a not uniform absorber (Turner et al. 1998,
Weaver et al. 1999) with denser layers 
near the equatorial plane (Matt et al. 2000). 
Accordingly to the model proposed for the Broad Line Region, one could 
think that the inner walls of the torus are responsible 
for the scattered light. In this case the scattering region should be 
completely ionized in order to act as a mirror.  
Although plausible, this interpretation can not completely explain 
the luminosity of the second component.
The ratio between direct and scattered radiation f=$n_2/n_1\sim 0.3$ is 
expressed as a function of the solid angle subtended by the 
scattering source and the Thompson depth f$=\tau_{sc} \times \Omega/4\pi$.
Even assuming that all the scattered light is seen by the observer through 
the thinner edge (i.e. that we have a direct view of the equatorial plane),
with a  column density  $\sim 10^{23}$ $cm^{-2}$,
the scattered radiation is less than 10$\%$.

Another appealing interpretation involves photoionized gas within the BLR.
In this case the scattered radiation should be observed through the
same gas/dust layer which obscures the BLR, 
assuming large dust grains in the circumnuclear AGN clouds (Maiolino et al
  2001a, 2001b) to reconcile the 
the discrepancy between 
the estimated $N_{H,2} \sim 7 \times 10^{22}$ and the observed 
low reddening of the BLR.

The idea that electron scattering can occur in BLR has been 
proposed in the past to explain the large wings observed in the
optical lines. Very recently, Laor (2007), analyzing high quality 
Keck observations of NGC4395, has shown that 
the H${\alpha}$  exponential wings 
can be actually produced within the BLR gas if $\tau_{sc}=0.34$.
However this argument is still controversial.
Although  it seems plausible that $\tau_{sc}\le 1$, 
the real value of the electron scattering depth is still very uncertain.

In the case of 3C 445, $\tau_{sc}\le 1$ requires  a solid angle subtended by 
the BLR $\Omega/4\pi \ge 0.3$ in order to deviate towards the observer 
$\sim 30\%$ of the primary radiation.
If the semi-aperture angle of the torus is equal to the inclination 
angle $i= 60^{\circ}$, the estimated solid 
angle  $\Omega/4\pi = 0.25$ is quite similar to 
that required to account for f=$n_2/n_1\sim 0.3$.
However, if $\tau_{sc}$ is significantly smaller,  
for example it is near to the value proposed by Laor (2007),  
the high ratio between direct and scattered radiation observed in
3C 445 requires a larger solid angle for the BLR or an
additional source of photons (for example, the inner wall of the torus).

\subsection{Soft X-ray spectrum}
Thanks to the unprecedented combination of energy resolution and
sensitivity of the RGS detector, the nature of the soft excess has
been resolved in 3C~445 for the first time. The RGS continuum exhibit
typical features of Seyfert~2 spectra.  Based on the Bianchi \&
Guainazzi (2007) diagnostic plane, we argue that the emission lines
originate in a warm gas photoionized by the nuclear engine.  The same
material is most likely responsible for scattering some of the nuclear
flow toward the observer. Indeed the ratio between the flux of the
scattered and primary power law is $2-3$ per cent, in very good
agreement with typical values observed in Seyfert 2 galaxies (Mulchaey
et al. 1993, Turner et al. 1997; Bianchi, private communication).
Since the ionized material is only attenuated by Galactic N$_H$, it is
natural to put it outside the torus and well beyond the BLR.  The
scattering/emitting gas could indeed coincide with the Narrow Line
Region, as recently suggested for Seyfert 2 galaxies by Bianchi,
Guainazzi \& Chiaberge (2006) on the base of the impressive
morphological similarity between Chandra and narrow-band HST images of
8 sources. 
Considering the strong similarity between Seyfert 2s and 3C 445 
(scattered continuum and emission lines), a non thermal origin of the 
soft X-ray continuum appears very improbable.
The contribution of an extra nuclear jet to the spectrum below 2 keV, if
present, has to be however marginal.

The detection of warm photo-ionized gas in the nuclear regions of Radio
Galaxies is, in some regards, unexpected. Since ASCA, it is well known that 
warm gas, producing characteristic absorption spectral features, is present 
in the inner region of Seyferts. 
On the contrary, up to now, the only cold
absorber has been unambiguously revealed in Radio Loud AGN.
This result is very important and needs to be consolidated by large sample 
analysis.  We note in passing that, if confirmed, it 
would support the (still debated) idea 
that soft X-ray deficit seen in a handful of high 
redshift (z$>4$) radio loud quasars is due to 
warm gas (Worsley et al. 2004, Yuan et al. 2005). 
It is indeed plausible that warm absorber seen in face-on sources
is the same material that scatters nuclear X-rays into our 
line of sight in sources at larger inclination angle.



\section{Summary and Conclusion}

\begin{itemize}

\item An XMM-Newton observation of the BLRG 3C 445
reveals its hidden Seyfert 2 nature. 

\item For the first time the soft excess of 3C 445 is resolved. It is well 
fitted with highly ionized lines over-imposed to a weak continuum.
Warm possibly outflowing gas located well beyond the torus, is responsible
for the observed features.  The same emitting line  gas deviates  
a fraction (3\%) of the primary X-ray continuum, escaping from the 
torus cone, towards the observer.

\item  The emerging picture is coherent and does not require an extra-nuclear
jet to explain the soft unabsorbed flux.

\item About $70\%$ of the nuclear hard continuum is dominated by 
an accretion flow radiation transmitted through a column density 
$\sim 4\times10^{23}$ cm$^{-2}$.  The detected $K_{\alpha}$ line has an 
equivalent width consistent with an origin in the line-of-sight absorber.
About $30\%$  of the nuclear emission is absorbed
by a more tenuous column density $\sim 7\times10^{22}$ cm$^{-2}$.

\item If the the observer has a deep view into the nucleus 
but near the edge of the torus, light scattered by the inner wall of 
the torus and/or  by the BLR region could (at least in part) account for 
the less absorbed nuclear component. 

\end{itemize}


{}

{\it Note} -- While we were waiting for the comments of the referee, 
Sambruna et al. (arXiv:0704.3053) submitted a similar paper to Astro-ph.
Our analysis is in agreement with their results.
\end{document}